\newcommand{\pam}{.\hskip-2pt$^\prime$}
\newcommand{\lsim}{\;\lower.6ex\hbox{$\sim$}\kern-7.75pt\raise.65ex\hbox{$<$}\;}
\newcommand{\degs}{$^{\circ}$}
\documentclass[11pt,twoside]{article}
\usepackage[dvips]{graphicx}
\usepackage{newpasp}
\begin{document}
\title{A Study of the H$_2$O Maser Emission from R~Cas}
 \author{J. Brand$^1$, L. Baldacci$^{1,2}$, D. Engels$^3$}
\affil{$^1$Istituto di Radioastronomia, CNR, Via Gobetti 101,
I-40129 Bologna, Italy}
\affil{$^2$Dip.to di Astron., Univ. Bologna, Via Ranzani 1, I-40127 Bologna,
Italy}
\affil{$^3$Hamburger Sternwarte, Gojensbergweg 112, D-21029 Hamburg, Germany}

\begin{abstract}
Over the past decade we observed a sample of late-type stars
(supergiants, semi-regular variables, OH/IR stars, and Mira's) 3--4 times per
year in the 1.3-cm line of H$_2$O. The observations were carried out 
with the Medicina 32-m and the Effelsberg 100-m telescopes. In addition, 
a sub-sample of these stars was observed at several epochs with the VLA. 
In our analysis we also use data from the literature, for instance on mass 
loss and optical variability. The aim is to investigate the properties of the 
circumstellar outflows and to elucidate the maser-pumping mechanism. In the
conference poster we presented some preliminary results of a 4-star sub-set 
of the data base; in the small space available here, we briefly describe the
monitoring program, and discuss the Mira-variable R~Cas in some detail.
\end{abstract}

\section{Introduction}

Maser emission from the 6$_{16}-5_{23}$ rotational transition of water at
22~GHz is a common feature in circumstellar shells; so far, it has been
detected in the envelopes of about 500 stars ($\delta >-$30\degs; Valdettaro et
al. 2001). The masers are strongly variable, sometimes in phase with the
luminosity variations of the central star, and sometimes erratically,
including spectacular flares.

\noindent
Because of the high excitation of the emitting level (650~K above ground) the
masers must be located in the circumstellar envelope fairly close to the star
(at distances of $\lsim 10^{15}$~cm), thus tracing the mass motions close to 
the regions where the mass loss starts. The high variability of the maser 
emission may therefore reflect the turbulent motions in these regions, 
indicating that the mass loss process is not a smooth outflow, but may occur
in blobs. The individual peaks in the spectra correspond to individual maser
spots in spatial maps, and these might be interpreted as such blobs, or
density enhancements.

\noindent
In 1990 we started an extensive observing program of several tens of
late-type stars, primarily with the Medicina 32-m and Effelsberg
100-m telescopes, supplemented by observations with the VLA, and the ISO 
satellite. The scientific aims of this monitoring 
campaign are to analyze the maser variability as a function of both time and
stellar parameters (optical/IR variability, mass loss rate, spectral type,
IR colours). In particular we attempt to:

\smallskip\noindent
{$\bullet$}\ determine typical sizes and numbers of maser components;
\hfill\break\noindent
{$\bullet$}\ estimate lifetimes of maser features and timescales for change;
\hfill\break\noindent
{$\bullet$}\ derive thickness and radial distance of the maser shell as a
function of stellar variations (energy output and mass-loss rate);
\hfill\break\noindent
{$\bullet$}\ probe the influence of the pumping conditions on the observed
maser properties.

\noindent
The campaign was most active in the period 1990-1995, but we have continued
single-dish observations of a sub-sample of stars up to the present day. In 
addition, for several of the stars we have (Medicina) observations between 
1987-1990, taken from the Arcetri archives.

\noindent
We have started the analysis of those stars for which we have a
significant time-coverage of the maser emission. As an example of the
available data, we here present preliminary results of one of these stars, 
R~Cas (Mira).
Fig.~1 gives a graphical representation of the behaviour of
the maser flux density versus velocity as a function of time.

\begin{figure*}[ht]
\resizebox{7cm}{!}{\includegraphics{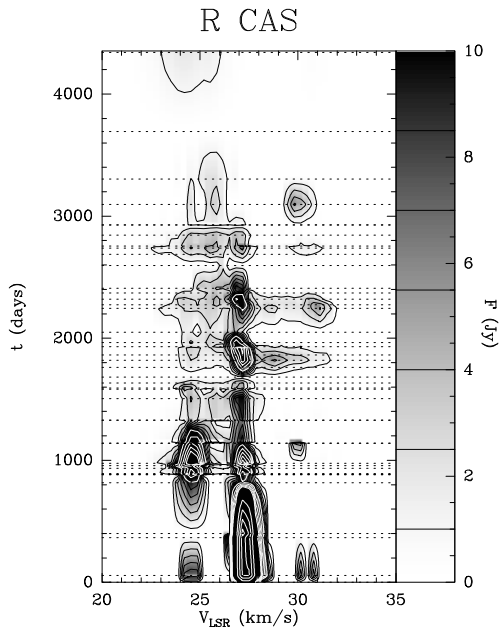}}
\parbox[b]{65mm}{
\caption{Grey-scale plots and contour map of the H$_2$O flux density versus
velocity as a function of time for R~Cas. The horizontal
dotted lines indicate when a spectrum is available; observations separated by
less than 10$^{\rm d}$ were averaged. Between adjacent observations linear
interpolation was performed. Contour values (start(step)end) are
1(1.5)8.5~Jy (black); 10(5)30, 100(50)300~Jy (white). The grey scale has
been adjusted to bring out the low-level emission. The velocity
resolution is 0.33~km\,s$^{-1}$}
}
\end{figure*}

\section{Observations and Results}

The observations were carried out with the Medicina 32-m and the
Effelsberg 100-m antennas. The Medicina telescope HPBW at 22.2~GHz is 1\pam9.
The pointing accuracy is better than $\sim 25^{\prime\prime}$.  We estimate
an uncertainty of $\sim$ 30$\%$ on the absolute flux scale, and a typical
1$\sigma$ rms noise of $\sim$1.5~Jy. The Effelsberg HPBW is
40$^{\prime\prime}$; the pointing accuracy is better than $\sim
8^{\prime\prime}$. Typical rms values are 0.2~Jy. 

\medskip\noindent
The observational results cover the period from 10 Sept. 1987 to 31 July 1999.
The time-velocity-flux density plot in Fig.~1 shows that the maser
emission is dominated by 2 components, at $V_{\rm lsr} \approx$24.5 and
27.1~km\,s$^{-1}$. This is further illustrated by the collection of spectra
shown in Fig.~2. 

\begin{figure}[ht]
\centerline{\resizebox{10cm}{!}{\rotatebox{270}{\includegraphics{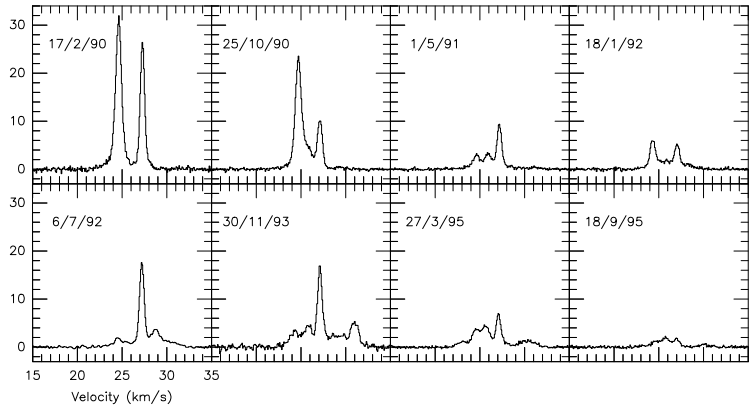}}} }
\caption{A collection of representative spectra from the R~Cas data base.
The vertical scale is in Jy; the date of observation is indicated in each
panel. The spectra are dominated by emission at $\sim$24.5 and
27.1~km\,s$^{-1}$; the remaining components are collectively called
``plateau'' in this paper.}
\end{figure}

\noindent
Maser emission has been detected most of the time, except on three occasions
when only (1$\sigma$, 0.33~km\,s$^{-1}$ resolution) upper limits of 2.9~Jy
(18/4/92) and 0.8~Jy (18/10/94, 15/9/95) could be determined.
Over the whole observing period, emission at levels $>$4~Jy (at resolution 
0.33~km\,s$^{-1}$) is found between 20.4 and 32.3~km\,s$^{-1}$.
With time, the maser strength has gradually decreased. This is seen very 
clearly in Fig.~3a, where we plot the integrated H$_2$O emission as a
function of time. The curve (least-squares fit for t$>$800$^{\rm d}$) is a
sinusoid with a period of $434 \pm 5^{\rm d}$ (equal to the optical period,
as derived from AAVSO data),
damped with an exponential (exp($\alpha$t), $\alpha=(-6.6\pm 1.0) \times
10^{-4}$).
The maser peaks with a $\sim$115$^{\rm d}$ delay with respect to the
optical maximum
output (indicated by dotted lines in the figure). The points indicated as
``flares'' were identified as such by comparison with data from Pashchenko
(1990), and were excluded from the fit. The Pashchenko data show 3 outbursts 
of the 27~km\,s$^{-1}$ component between mid-1986 and mid-1989, with the flux 
density reaching values of 600~Jy in early 1989, with which our data are
consistent (note however, that
including these points in the fit changes the exponent of the damping
function, but does not significantly affect the derived period).
The overall decrease of the 
integrated flux density with time is due to the decrease in flux
density of the 2 main components: removing their contribution from the total
flux density leaves the emission in the other, less intense components that
cluster around the main lines (called ``the plateau''). This latter
component, shown in Fig.~3b, varies with the same period, 
and does not show the damping effect seen in the total integrated flux density.

\begin{figure}[ht]
\centerline{
\resizebox{6.5cm}{!}{\rotatebox{270}{\includegraphics{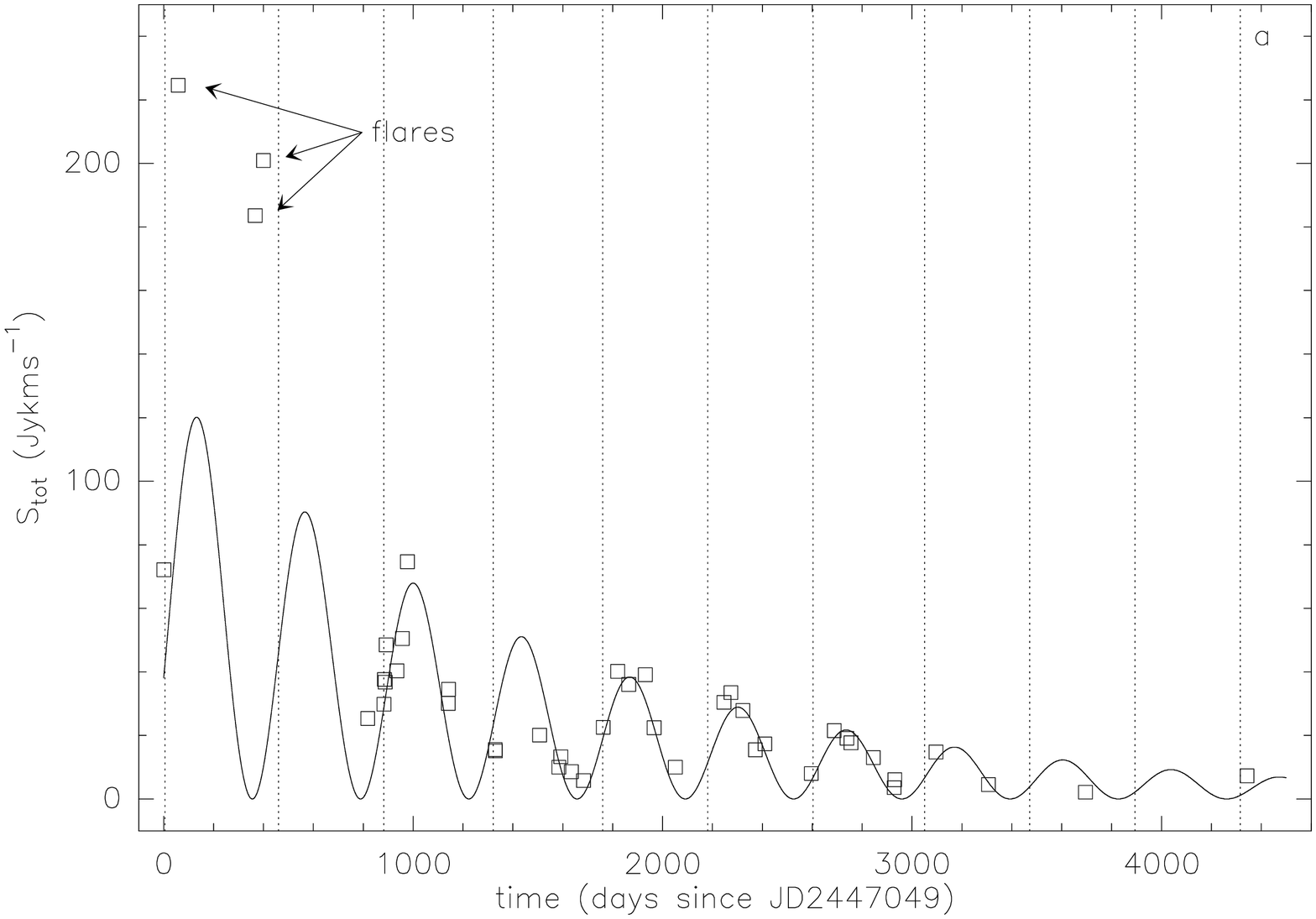}}} 
\hspace{1cm}
\resizebox{6.5cm}{!}{\rotatebox{270}{\includegraphics{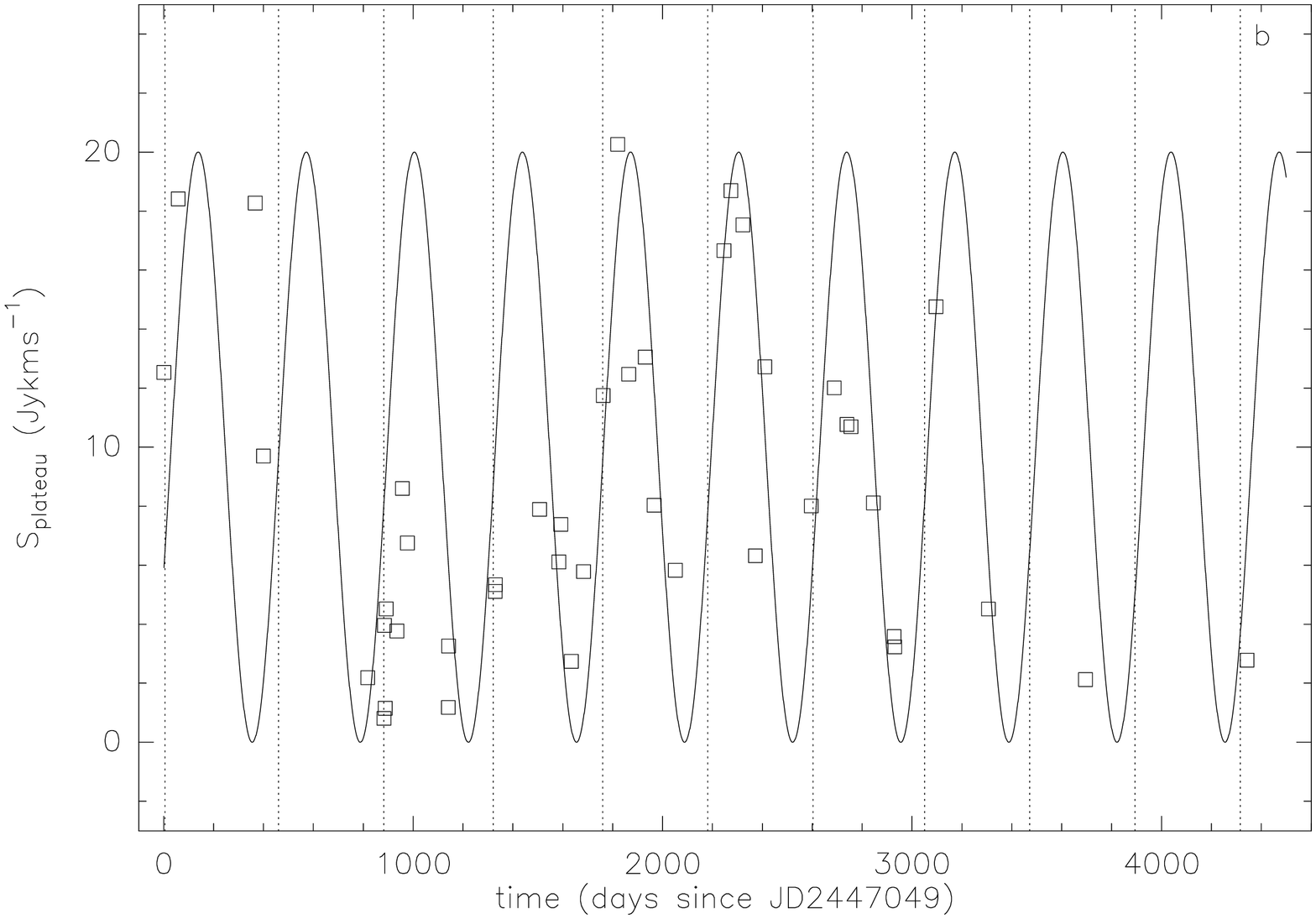}}}
}
\caption{{\bf a}\ Integrated (10--40~km\,s$^{-1}$) H$_2$O emission of R~Cas 
versus time. Day=0 corresponds to JD2447049. The squares are the data, the 
curve is a fit of an exponentially damped sinusoid with a period of
$434 \pm 5^{\rm d}$. The vertical dotted lines mark the times of maximum
optical output (AAVSO data), with respect to which the maser maximum has a 
delay of $\sim$115$^{\rm d}$. {\bf b}\ As a, but for the integrated emission 
of the ``plateau'' (total emission minus the contribution of the main 
components at 24.5 and 27.1~km\,s$^{-1}$. The curve shown has P=433$^{\rm d}$.}
\end{figure}

\section{Discussion and conclusions}

The H$_2$O maser emission from the circumstellar envelope of R~Cas might be
described as a combination of a ``plateau'', which consists of numerous
lines at a level of a few Jy, and two strong components (at $\approx$24.5 and
27.1~km\,s$^{-1}$) which dominated the emission for several years. The
``plateau'' varies regularly with the period of the star, displaced in phase
by $\sim$115~days. The superposition of ``plateau'' and strong components is
well described in terms of a damped oscillator. The lines of the ``plateau''
probably originate in randomly distributed `blobs' within the spherically
symmetric wind. The strong components occasionally flared and decreased
steadily since 1990; they might have traced two regions of enhanced density,
which have left the H$_2$O maser shell recently. With a size of 3.6$\times
10^{14}$~cm (Colomer et al. 2000) the typical crossing time of the R~Cas
maser shell is about 30~years, and the past decade we may have witnessed 
the passage of a density inhomogeneity through the shell. 
As recent as May 2001 the total maser output was still at a very low level; it 
will be interesting to see if and when new strong components will appear 
again. 

\acknowledgements
In this research, we have used, and acknowledge with thanks, data
from the AAVSO International Database, based on observations submitted to
the AAVSO by variable-star observers worldwide. We thank Gianni Comoretto
and Riccardo Valdettaro for presenting the poster paper at the meeting in
our absence.

\begin{center} {\bf References} \end{center}

\noindent Colomer F., Reid M.J., Menten K.M., Bujarrabal V., 2000, A\&A 355,
979

\noindent Pashchenko M.I., 1990, Astron. Tsirk 1543

\noindent Valdettaro, R., Palla, F., Brand, J. et al. 2001, A\&A, 368, 845 

\end{document}